\documentclass[letterpaper]{article} 
\usepackage{aaai2027}  
\usepackage[hyphens]{url}  
\usepackage{graphicx} 
\usepackage{natbib}  
\usepackage{caption} 
\usepackage{algorithm}
\usepackage{algorithmic}

\usepackage{newfloat}
\usepackage{listings}
\DeclareCaptionStyle{ruled}{labelfont=normalfont,labelsep=colon,strut=off} 
\floatstyle{ruled}
\newfloat{listing}{tb}{lst}{}
\floatname{listing}{Listing}

\usepackage{booktabs}

\usepackage{multirow} 

\usepackage{amsmath} 
\usepackage{amssymb}

\usepackage[most]{tcolorbox}
\tcbuselibrary{listings,breakable}

\newtcblisting{promptbox}{
    enhanced,
    listing only,
    breakable,
    colback=white,
    colframe=black,
    boxrule=0.8pt,
    arc=0pt,
    outer arc=0pt,
    left=1.5mm,
    right=1.5mm,
    top=1.2mm,
    bottom=1.2mm,
    listing options={
        basicstyle=\ttfamily\footnotesize,
        breaklines=true,
        columns=fullflexible,
        keepspaces=true,
        showstringspaces=false,
        tabsize=2
    }
}
\newtcolorbox{examplebox}[1]{
    enhanced,
    breakable,
    colback=white,
    colframe=black,
    boxrule=0.8pt,
    arc=0pt,
    outer arc=0pt,
    left=1.5mm,
    right=1.5mm,
    top=1.2mm,
    bottom=1.2mm,
    title=#1,
    fonttitle=\bfseries
}

\title{AudioDER: A Deduplication-Enhanced Reasoning Dataset for Post-Training Large Audio-Language Models}
\author{
    Geng Hui\textsuperscript{\rm 1}\textsuperscript{\rm 3}\equalcontrib,
    Yi Su\textsuperscript{\rm 1}\textsuperscript{\rm 2}\equalcontrib,
    Zijian Gao\textsuperscript{\rm 1}\textsuperscript{\rm 3},
    Tianjiao Wan\textsuperscript{\rm 1}\textsuperscript{\rm 3},
    Qisheng Xu\textsuperscript{\rm 1}\textsuperscript{\rm 2},
    Jiaxin Chen\textsuperscript{\rm 1}\textsuperscript{\rm 3},
    Hengzhu Liu\textsuperscript{\rm 1},
    Kele Xu\textsuperscript{\rm 1}\textsuperscript{\rm 3}\corresponding
}
\affiliations{
    \textsuperscript{\rm 1} College of Computer Science and Technology, National University of Defense Technology\\
    \textsuperscript{\rm 2}National Key Laboratory of Parallel and Distributed Computing, National University of Defense Technology\\
    \textsuperscript{\rm 3}State Key Laboratory of Complex \& Critical Software Environment, National University of Defense Technology\\
    Changsha, Hunan 410000 China
}

\begin{document}

\maketitle

\begin{abstract}
Recent advances in pretrained large audio-language models (LALMs) have demonstrated strong capabilities across speech, sound, and music. To adapt these models to downstream tasks without the cost of pretraining from scratch, post-training has become a widely adopted paradigm. However, the effectiveness of post-training depends critically on the quality of the training corpus. 
We observe that existing post-training corpora, often constructed by aggregating public audio datasets, suffer from substantial acoustic redundancy, as many of these datasets are sourced from overlapping media platforms. Such redundancy leads to repeated exposure to similar acoustic patterns, causing diminishing returns in performance despite increased data volume. 
address this issue, we propose a three-stage data construction pipeline that performs acoustic redundancy filtering, converts retained samples into a unified multiple-choice question-answering format with chain-of-thought generation, and finally applies quality verification and filtering. Using this pipeline, we construct AudioRE, a post-training dataset of approximately 286k instances spanning sound, speech, and music. Supervised fine-tuning on AudioRE consistently improves the performance of Qwen2-Audio-7B-Instruct across diverse audio understanding and reasoning benchmarks, outperforming models trained on the unfiltered raw corpus with substantially more instances. 
These results validate the effectiveness of our redundancy-aware data construction pipeline and the resulting AudioRE dataset, and further highlight the importance of minimizing acoustic redundancy in audio-language post-training. To facilitate future research, we will release both the AudioRE and the fine-tuned Qwen2-AudioRE checkpoint.
\end{abstract}


\begin{figure}[ht!]
\centering
\includegraphics[width=0.99\linewidth]{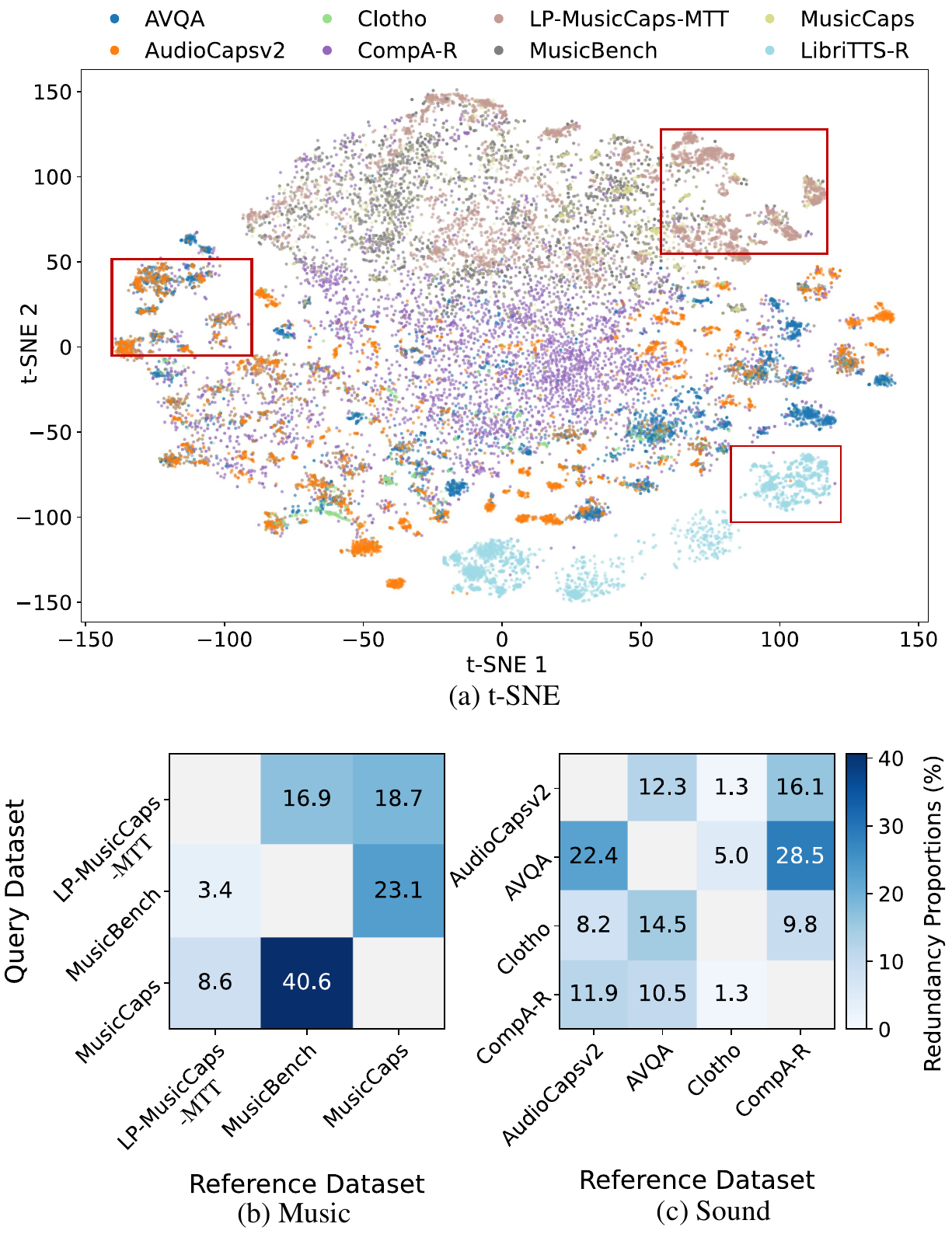}
\caption{
Analysis of acoustic redundancy.
\textbf{(a)} t-SNE visualization of audio clips from different source datasets, showing substantial overlap across datasets.
\textbf{(b)-(c)} Cross-dataset redundancy matrices for the music and sound domains, respectively. Each entry $R_{i,j}$ denotes the proportion of samples in query dataset $i$ that are involved in at least one redundant pair (cosine similarity exceeding 0.90) in the reference dataset $j$, with rows representing query datasets and columns representing reference datasets.
}
\label{fig:cross_dataset_similarity}
\vspace{-0.5cm}
\end{figure}


\section{Introduction}
Recent years have witnessed rapid progress in large audio-language models (LALMs). Through large-scale pretraining on audio-text corpora, models such as Qwen2-Audio~\cite{chu2024qwen2}, Audio Flamingo 3~\cite{goel2025audio}, Qwen2.5-Omni~\cite{xu2025qwen25omni} and MOSS-Audio~\cite{yang2026moss} have demonstrated broad foundational capabilities across speech, sound, and music, covering a wide range of audio understanding tasks.
As increasingly capable pretrained LALMs become publicly available, post-training has emerged as a practical paradigm for adapting these foundation models to downstream applications, bypassing the prohibitive cost of pretraining from scratch. 
However, the design of effective post-training corpora remains underexplored. Current construction efforts tend to emphasize scaling up dataset size, operating under the assumption that larger corpora yield stronger fine-tuned models. In practice, however, many audio-language datasets are sourced from a limited set of public repositories, and their direct aggregation introduces substantial acoustic redundancy. As a result, the model is repeatedly exposed to similar acoustic patterns, leading to diminishing returns in performance despite increased data volume.

To empirically verify this phenomenon, we investigate the acoustic overlap across the source datasets considered in this work. Figure~\ref{fig:cross_dataset_similarity}(a) presents a t-SNE visualization of embeddings extracted from Contrastive Language-Audio Pretraining (CLAP)~\cite{laion_clap2022} for audio samples drawn from all datasets. Clips from different datasets form tightly overlapping clusters and are intermingled across multiple regions of the embedding space, indicating that cross-dataset acoustic redundancy is widespread. To further quantify this overlap and illustrate its severity, we focus on the sound and music domains, where the redundancy is particularly pronounced. Specifically, we compute pairwise cosine similarities between every pair of audio clips across different datasets. Pairs with similarity exceeding 0.90 are considered redundant. For each ordered pair of datasets $(i, j)$, we define $R_{i,j}$ as the proportion of all samples in dataset $i$ that are involved in at least one redundant pair with dataset $j$. In this redundancy matrix, the row index $i$ denotes the query dataset and the column index $j$ denotes the reference dataset. Figures~\ref{fig:cross_dataset_similarity}(b) and (c) present these matrices for the music and sound domains, respectively. These results confirm that a substantial portion of samples in each query dataset have near-duplicates in the reference datasets, suggesting that simply aggregating existing public corpora leads to considerable repetitive exposure to similar acoustic patterns during fine-tuning.

These observations highlight a gap in existing post-training corpus construction: while dataset overlap and data leakage have received increasing attention~\cite{wijngaard2025audio}, acoustic redundancy has received little consideration as an explicit data-selection criterion, let alone been adopted as a guiding principle for corpus construction. This motivates our central research question: \emph{How can we construct an audio-language post-training corpus from public datasets that effectively reduces redundancy while preserving broad data coverage?}


To address this question, we propose a three-stage, redundancy-aware data construction pipeline. First, we extract audio embeddings using CLAP~\cite{laion_clap2022} and compute pairwise cosine similarities to identify and filter out acoustically redundant samples, retaining a diverse subset for subsequent annotation. Second, to support supervised fine-tuning (SFT) across diverse datasets in a unified format, we convert all retained samples and their original annotations into consistent multiple-choice question answerings (MCQAs), leveraging their original annotations, and use Qwen3-30B~\cite{yang2025qwen3} to generate a chain-of-thought (CoT) for each instance. Third, we apply quality filtering to ensure question clarity, option distinguishability, answer consistency, and CoT logical coherence. We apply this pipeline to several publicly available audio-language datasets, producing AudioRE, a post-training corpus containing approximately 286k instances across sound, speech, and music. Each instance consists of an audio clip, an audio caption, a four-option multiple-choice question, the correct answer, and a generated CoT. To validate the effectiveness of our pipeline, we post-train Qwen2-Audio-7B-Instruct on AudioRE and evaluate its performance across diverse audio understanding benchmarks. The results demonstrate that our redundancy-aware data construction approach consistently improves model capabilities, underscoring the importance of acoustic diversity in post-training data curation.
Our main contributions are summarized as follows:
\begin{itemize}
\item We identify and systematically analyze acoustic redundancy arising from the aggregation of different public audio datasets.
\item We propose a three-stage data construction pipeline that integrates acoustic redundancy filtering, unified MCQA transformation with CoT generation, and quality filtering.
\item We construct and release AudioRE, a post-training corpus of approximately 286k instances across sound, speech, and music, along with Qwen2-AudioRE, a Qwen2-Audio-7B-Instruct checkpoint post-trained on the corpus.
\item We demonstrate that Qwen2-AudioRE consistently improves performance across diverse audio understanding benchmarks, outperforming existing post-training models built on the same backbone. Moreover, AudioRE achieves better results than our unfiltered raw corpus, despite containing substantially fewer instances.
\end{itemize}


\begin{table*}[t]
\centering
\small
\begin{tabular}{ccccc}
\toprule
\textbf{Dataset}& \textbf{Domain}& \textbf{Annotation Origin}
& \textbf{\# Instances}& \textbf{Media Source} \\
\midrule
AVQA& Sound& Human& 39,966& YouTube \\
Clotho& Sound& Human& 2,865& Freesound \\
AudioCaps v2& Sound& Human& 90,549& YouTube \\
CompA-R& Sound& Pipeline& 197,218& YouTube \\
MusicCaps& Music& Human& 2,608& YouTube \\
MusicBench& Music& Pipeline&17,861&YouTube \\
LP-MusicCaps-MTT& Music& Pipeline& 15,560& MagnaTagATune \\
LibriTTS-R& Speech& Pipeline& 226,183& LibriVox \\
\bottomrule
\end{tabular}
\caption{
Overview of source datasets used to construct AudioRE.
}
\label{tab:source_datasets}
\end{table*}

\begin{table*}[t]
\centering
\small
\begin{tabular}{lccccccc}
\toprule
\textbf{Dataset} & \textbf{Scale} & \textbf{Caption} & \textbf{MCQA} & \textbf{CoT} &\textbf{\# Sources}& \textbf{Acoustic Redundancy Filtering} & \textbf{Availability} \\
\midrule
CoTA~\cite{xie2025audio}  & 1.2M & \checkmark & N/R & \checkmark&9 & N/R & Annotations only \\
AudioMCQ~\cite{he2026measuring} & 571k & \checkmark & \checkmark & \checkmark&7 & N/R & \checkmark \\
SARI~\cite{wen2025sari} & 32k & \checkmark & \checkmark & \checkmark&4 & N/R & N/R \\
\midrule
AudioRE (ours) & 286k & \checkmark & \checkmark & \checkmark &8& \checkmark & \checkmark \\
\bottomrule
\end{tabular}
\caption{Comparison of AudioRE with other post-training datasets for LALMs.}
\label{tab:posttraining_dataset_comparison}
\vspace{-0.5cm}
\end{table*}

\section{Related Work}\label{related_work}
\textbf{Large Audio Language Models.}
Large audio-language models (LALMs) connect audio perception with language understanding and generation, providing a unified interface for tasks involving speech, sound, and music. A common design couples a pretrained audio encoder to a large language model through a modality adapter. Representative systems include LTU~\cite{gong2024listen}, LTU-AS~\cite{gong2023joint}, Pengi~\cite{deshmukh2023pengi}, SALMONN~\cite{tang2024salmonn}, GAMA~\cite{ghosh2024gama}, and Qwen2-Audio~\cite{chu2024qwen2}, which demonstrate rapid progress in general-purpose audio understanding. The Audio Flamingo family has evolved from few-shot learning and multi-turn dialogue to long-audio understanding and expert reasoning, and then to reasoning, multi-audio dialogue, and voice interaction across speech, sound, and music~\cite{kong2024audio,ghosh2025audio,goel2025audio}. Audio Flamingo Next further extends long-context and temporally grounded audio reasoning~\cite{ghosh2026audio}. Kimi-Audio~\cite{ding2025kimi} supports audio understanding, generation, and conversation, whereas Moshi~\cite{défossez2024moshi} targets real-time full-duplex spoken dialogue by modeling user and system speech in parallel streams. MOSS-Audio~\cite{yang2026moss} is a unified audio-language model for speech, environmental sound, and music understanding, supporting audio captioning, time-aware question answering, timestamped transcription, and audio-grounded reasoning.

\textbf{Post-Training.}
Improving the reasoning capabilities of LALMs has recently emerged as an important research direction. Audio-CoT~\cite{ma2025audio} systematically evaluates CoT prompting across speech, sound, and music, reporting clear gains on easy and medium-difficulty tasks but less reliable improvements on hard tasks.
Recent work also studies supervised fine-tuning (SFT) and reinforcement learning (RL) for audio reasoning. Audio-Reasoner~\cite{xie2025audio} constructs CoTA, a 1.2M-instance corpus with structured reasoning traces organized into planning, captioning, reasoning, and summarization.
R1-AQA~\cite{li2025reinforcement} applies SFT and group relative policy optimization (GRPO) to Qwen2-Audio-7B-Instruct and analyze the different prompts. SARI~\cite{wen2025sari} uses SFT warm-up with structured and unstructured CoT variants, followed by curriculum-guided GRPO on a 32k MCQ corpus.
Omni-R1~\cite{rouditchenko2025omni} fine-tunes Qwen2.5-Omni with GRPO and examines whether post-training gains arise from audio-conditioned learning or improved text-based reasoning.

\textbf{Datasets.}
Progress in audio reasoning has been supported by datasets containing captions, question-answering pairs, MCQAs, and CoT. Although originally introduced for audio-visual question answering, AVQA~\cite{yang2022avqa} provides human-annotated four-option questions involving audio-visual relationships in real-world videos. CoTA~\cite{xie2025audio} provides large-scale structured reasoning supervision across diverse audio tasks. SARI~\cite{wen2025sari} contributes 32k MCQ samples with structured and unstructured CoT supervision. AudioMCQ~\cite{he2026measuring} contains approximately 571k MCQ samples with structured and unstructured CoT annotations and introduces audio-contribution filtering to distinguish weak- and strong-audio-contribution examples. 
AudioRE constructs a lower-redundancy corpus while preserving the available annotations of retained samples in a unified format.

\begin{figure*}[t]
\centering
\includegraphics[width=0.99\linewidth]{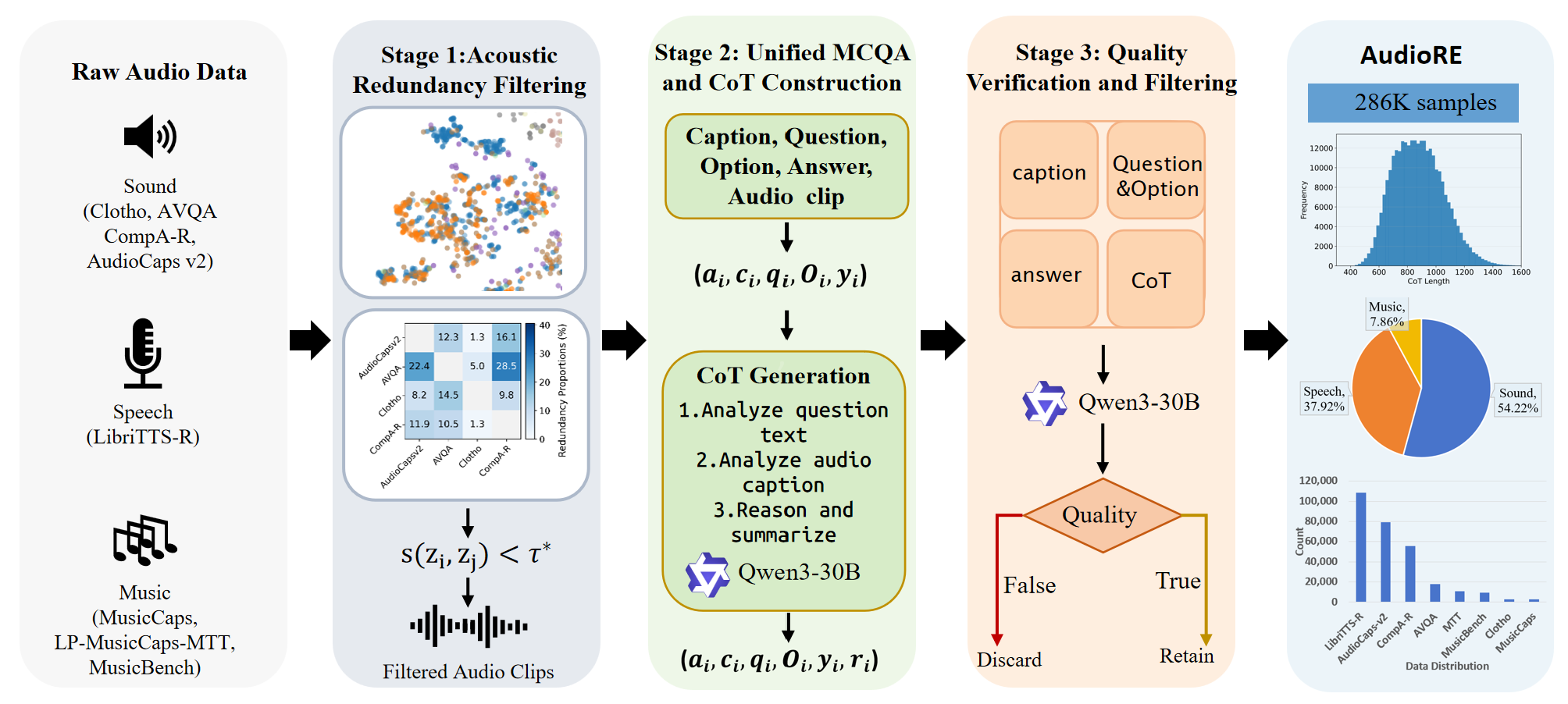}
\caption{Overview of the AudioRE construction pipeline.
Starting from eight audio datasets spanning sound, speech, and music, we first perform CLAP-based acoustic redundancy analysis and filtering. We then normalize the collected captions and question-answer annotations into a unified four-option MCQ format and generate CoTs for the retained samples. Finally, we apply quality filtering to obtain reliable training instances.}
\label{AudioRE_pipeline}
\end{figure*}

\section{AudioRE Dataset}\label{AudioRE_dataset}
\subsection{Overview}
AudioRE is constructed from eight publicly available audio datasets covering three major domains: sound, speech, and music. These datasets include AVQA~\cite{yang2022avqa}, Clotho~\cite{drossos2020clotho}, AudioCaps v2~\cite{kimAudioCaps2019}, CompA-R~\cite{ghosh2024gama}, MusicCaps~\cite{agostinelli2023musiccaps}, MusicBench~\cite{melechovsky2024mustango}, LP-MusicCaps-MTT~\cite{doh2023lp},
and LibriTTS-R~\cite{koizumi2023libritts}. Table~\ref{tab:source_datasets} summarizes their audio domains, annotation origins, scales, and primary media sources. More details are provided in the supplementary material. 
Table~\ref{tab:posttraining_dataset_comparison} provides a descriptive comparison between AudioRE and representative post-training datasets for LALMs. AudioRE differs from existing corpora in two key aspects. First, it explicitly incorporates acoustic redundancy analysis and filtering as a core design principle during corpus construction, mitigating the issue of repetitive exposure to similar acoustic patterns. Second, it enriches each audio clip with multiple forms of annotation, including a descriptive caption, a MCQA, and a CoT, providing comprehensive supervision signals for post-training. The construction process consists of three stages: (1) Acoustic Redundancy Analysis and Filtering, (2) Unified MCQA and CoT Construction, and (3) Quality Verification and Filtering. The overall construction pipeline is illustrated in Figure~\ref{AudioRE_pipeline}.

\subsection{Data Construction}
\textbf{Stage 1: Acoustic Redundancy Filtering.}

Based on the observed cross-dataset acoustic redundancy and the fact that clips from the same domain tend to exhibit higher similarity than cross-domain clips, we perform domain-specific filtering for sound, music, and speech. The filtering procedure consists of preprocessing, similarity computation, and Pareto-based heuristic sample selection.

Specifically, we first retain clips with durations between 5 and 30 seconds, and resample each waveform to 48 kHz. We then extract an audio embedding using the audio encoder of the \texttt{larger\_clap\_general} checkpoint from LAION-CLAP~\cite{laion_clap2022}.

Let $\widetilde{\mathbf z}_i=\operatorname{CLAP}(a_i)$ denote the embedding of audio clip $a_i$, and let $\mathbf z_i=\frac{\widetilde{\mathbf z}_i}{\left\|\widetilde{\mathbf z}_i\right\|_2}$ denote its $\ell_2$-normalized representation. We use CLAP embedding similarity as acoustic redundancy. The acoustic similarity between two clips $a_i$ and $a_j$ is then defined as $s_{ij}=\mathbf z_i^\top\mathbf z_j$.

To select a practical threshold \(\tau_m\) for each domain \(m\), we adopt a Pareto-based heuristic that balances two inherently conflicting objectives: minimizing corpus coverage loss and minimizing local redundancy among retained samples. We define these two objectives below and then apply Pareto optimization to identify the threshold that achieves the best trade-off.
Let \(\mathcal A_m\) denote the set of duration-valid candidate clips in domain \(m\). A candidate clip is retained only if its maximum similarity to all previously retained clips is below \(\tau\), where the maximum over an empty set is defined as \(-\infty\). The final retained subset is denoted by \(\mathcal K_m^{(\tau)}\subseteq\mathcal A_m\). Specifically, source datasets are first grouped according to annotation quality, with human-annotated datasets assigned higher priority than pipeline-generated datasets. Clips are then interleaved across source datasets using a deterministic schedule. This prioritization is intended to preserve annotations that are more directly derived from human supervision. We define the coverage distance:
\begin{align}
C_m(\tau)=\frac{1}{|\mathcal A_m|}\sum_{a_i\in\mathcal A_m}\left[1-
\max_{a_j\in\mathcal K_m^{(\tau)}}
\mathbf z_i^\top\mathbf z_j
\right].
\end{align}
A lower \(C_m(\tau)\) indicates that the corpus remains well represented by the retained set. 

To measure local redundancy, we compute the nearest-neighbor distance for every retained clip and define \(D_m(\tau)\) as:
\begin{align}
d_i^{\mathrm{NN}}(\tau)=1-\max_{\substack{a_j\in\mathcal K_m^{(\tau)}\\j\neq i}}
\mathbf z_i^\top\mathbf z_j.
\end{align}
\begin{align}
D_m(\tau)
\triangleq
Q_{0.10}
\left(
\left\{
d_i^{\mathrm{NN}}(\tau):
a_i\in\mathcal K_m^{(\tau)}
\right\}
\right),
\end{align}
where \(Q_{0.10}\) denotes the 10th percentile. We use the 10th percentile because it characterizes the lower tail of the nearest-neighbor distance distribution, where high-similarity clusters are concentrated. Compared with the minimum, it is less sensitive to a single anomalous pair, while unlike the mean or median, it remains sensitive to the most crowded portion of the retained embedding space. It therefore provides a practical balance between sensitivity to local redundancy and statistical stability. The nearest-neighbor distance is computed for retained subsets containing at least two clips. A larger \(D_m(\tau)\) indicates greater separation among clips in the locally crowded portion of the retained embedding space. Both \(C_m(\tau)\) and \(D_m(\tau)\) are computed exactly using blockwise matrix multiplication.

Each candidate threshold is represented by the two-objective point $\left(C_m(\tau), D_m(\tau)\right)$, where coverage distance $C_m(\tau)$ is minimized and nearest-neighbor distance $D_m(\tau)$ is maximized. After removing dominated thresholds, we independently apply min-max normalization to the coverage distance and nearest-neighbor distance over the Pareto-optimal threshold set. The normalized values are denoted by \(\overline C_m(\tau)\) and \(\overline D_m(\tau)\), respectively. The coverage-oriented endpoint is the Pareto-optimal threshold with the smallest normalized coverage distance, whereas the separation-oriented endpoint is the Pareto-optimal threshold with the largest normalized nearest-neighbor distance. We define \(\mathcal P_m\) as the set of the remaining Pareto-optimal thresholds after excluding these two endpoints. We select
\begin{align}
\tau_m^{\star}
\in \operatorname*{arg\,max}_{\tau\in\mathcal P_m}
\operatorname{dist}_{2}
\left(
\left[
\overline C_m(\tau),
\overline D_m(\tau)
\right],
\ell_m
\right),
\end{align}
where \(\ell_m\) denotes the line segment connecting the normalized objective coordinates of the coverage-oriented and separation-oriented endpoints and \(\operatorname{dist}_{2}(\mathbf x,\ell_m)\) denotes the Euclidean distance from point \(\mathbf x\) to the line segment \(\ell_m\). For all three domains, the Pareto-optimal set contains at least one non-endpoint candidate. More details are provided in the supplementary material.

\textbf{Stage 2: Unified MCQA and CoT Construction.}

After acoustic redundancy filtering, all retained audio clips are converted into training instances in a unified multiple-choice question-answering format. We preserve and adapt the available annotations from each source dataset according to its original annotation type; detailed provenance and processing operations are provided in the supplementary material. For example, AVQA~\cite{li2025reinforcement} is originally designed for audio-visual question answering. Since our setting is audio-only, we normalize explicit modality references in the question wording from ``video'' to ``audio'', retain the original options and answers, and filter out visual-dependent questions during quality control. For each audio clip, we use its original descriptive caption as a prompt to generate the corresponding CoT.

Each unified MCQA is represented as $x_i=(a_i,c_i,q_i,O_i,y_i)$, where $a_i$ denotes the audio clip, $c_i$ is its audio caption, $q_i$ is the question, $O_i=(o_i^1,o_i^2,o_i^3,o_i^4)$ is a tuple of four answer
options, and $y_i$ is the unique correct answer. Following prior work~\cite{xie2025audio,wen2025sari}, we use Qwen3-30B~\cite{yang2025qwen3} to generate a CoT \(r_i\) for each MCQA sample, conditioned on the source caption, question, answer options, and the annotated correct answer: $r_i=G_{\theta}(c_i,q_i,O_i,y_i)$. 
The generation prompt instructs Qwen3-30B to identify the information required by the question, extract the relevant audio and semantic evidence described in the source caption, and connect this evidence to the answer options. The prompt further requires the generated CoT to remain faithful to the source annotations and avoid introducing unsupported audio events, attributes, or inferences. After CoT generation, each training instance is represented as $\widetilde{x}_i=(a_i,c_i,q_i,O_i,y_i,r_i)$. 

\textbf{Stage 3: Quality Verification and Filtering.}
Finally, we apply quality verification and filtering to remove obvious inconsistencies and improve dataset reliability. For each candidate instance $\widetilde{x}_i$, we provide Qwen3-30B with a review prompt containing the audio caption, the question, the answer options, the annotated correct answer, and the generated CoT.
Qwen3-30B examines whether the annotation fields form a valid and consistent multiple-choice instance. Specifically, it checks whether the question is clear, whether the four answer options are mutually distinguishable and include a unique correct answer, and whether the annotated answer is supported by the audio caption. It further assesses the logical coherence of the CoT, checking obvious inconsistencies between the CoT and available annotations.

The quality-review model produces a binary decision $v_i=F_{\phi}(c_i,q_i,O_i,y_i,r_i)\in \{\texttt{True},\texttt{False}\}$. An instance is retained only when it receives a \texttt{True} decision. Instances receiving a \texttt{False} decision are discarded. This filtering step reduces answer--annotation inconsistencies, logically incoherent CoTs, and content unsupported by the available source annotations. In addition, we conducted a manual spot-check on the filtered data. The results confirm that the quality filtering effectively removed inconsistent instances, and the remaining samples exhibit satisfactory quality in terms of questions, options, and CoTs, validating the effectiveness of this step.

\begin{figure}
\centering
\includegraphics[width=0.99\linewidth]{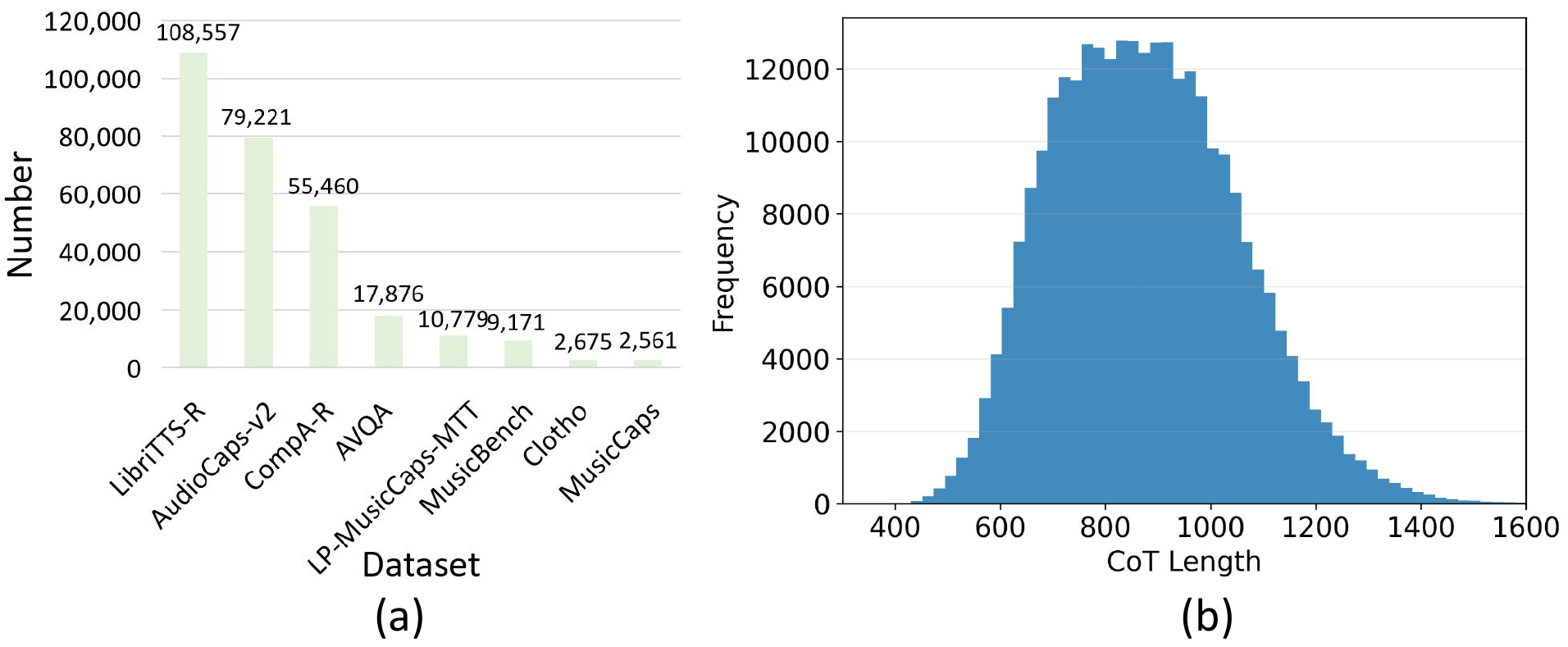}
\caption{ The Distribution analysis of AudioRE dataset.
(a) Distribution of the final AudioRE instances across the eight source datasets. (b) Distribution of CoT lengths in AudioRE. 
}
\label{data_num}
\end{figure}

\begin{figure}[t]
\centering
\includegraphics[width=0.94\linewidth]{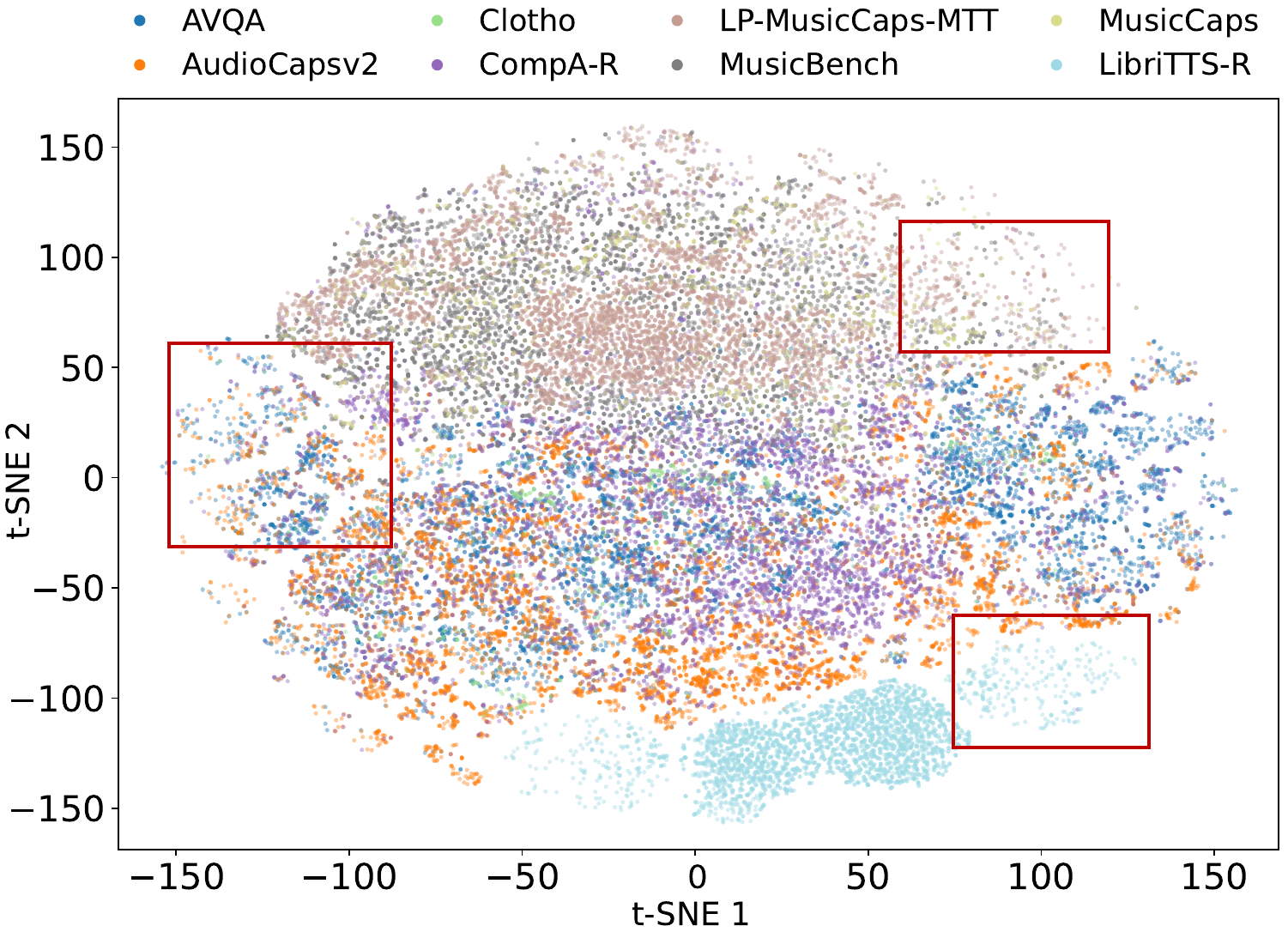}
\caption{
t-SNE visualization of CLAP embeddings of audio clips in AudioRE. Compared with the original data in Figure~\ref{fig:cross_dataset_similarity}(a), the retained samples are more evenly distributed with fewer high-density regions, reflecting the effect of redundancy filtering.
}
\label{after_tsne}
\end{figure}

\begin{table*}[t]
\centering
\small
\begin{tabular}{c|l|cccc|c|c}
\toprule
\multirow{2}{*}[-4pt]{\textbf{Model}}
&\multirow{2}{*}[-4pt]{\textbf{Method (dataset)}}
& \multicolumn{4}{c|}{\textbf{MMAU Test-mini}}
& \multicolumn{1}{c|}{\textbf{MMSU}}
& \multicolumn{1}{c}{\textbf{MMAR}} \\
\cmidrule{3-8}
&& Sound & Music & Speech & Average & Acc.& Acc. \\
\midrule
\multicolumn{8}{c}{
\textbf{Large Audio Language Models}
} \\
\midrule

LTU~\cite{gong2024listen}&Direct Inference &36.34 &29.94& 20.72 &29.00 
                   &  29.76  
                  &  23.30  \\
LTU-AS~\cite{gong2023joint}&Direct Inference &  23.35&  9.10&  20.60 & 17.68& - & 20.00
                  \\
SALMONN~\cite{tang2024salmonn}&Direct Inference & 41.14 & 37.13 &26.43&34.90
                   &  -
                  &  32.80 \\

Qwen-Audio-Chat~\cite{chu2023qwen}&Direct Inference & 55.25& 44.00& 30.03& 43.10&  -& 23.50 \\   
GAMA~\cite{ghosh2024gama}&Direct Inference &  41.44 & 32.33 & 18.91&30.90
                 &  34.44
                  & 27.70  \\ 
GAMA-IT~\cite{ghosh2024gama}&Direct Inference&43.24& 28.44& 18.91& 30.20&-&22.40\\
Mellow~\cite{deshmukh2025mellow}&Direct Inference & 61.26 & 54.19 & 29.73&48.40
                 &  -
                  & - \\  
Kimi-Audio~\cite{ding2025kimi}&Direct Inference & 54.35 & 55.39 & 54.95&54.90
                 &  56.64
                  & 47.00 \\            
\midrule
\multicolumn{8}{c}{
\textbf{Qwen2-Audio-7B-Instruct-based}
} \\
\midrule
Qwen2-Audio-7B-Instruct~\cite{chu2024qwen2}&Direct Inference & 67.27 & 56.29 & 55.26 &59.60
               &35.72 
                  & 30.00 \\
Audio-Reasoner~\cite{xie2025audio}& SFT (CoTA) &  60.06& 64.30& 60.70& 61.71
                  & 49.20
                  & 36.80 \\
SARI~\cite{wen2025sari}& SFT &  64.87&  59.13&  44.52&  56.18 
& 37.29 & - \\
R1-AQA~\cite{li2025reinforcement}&SFT (AVQA) &  60.96& 49.19& 45.35& 51.80
                     & 35.36
                  & -  \\                                       
Qwen2-AudioRE (Ours) &SFT (AudioRE)& \textbf{73.87} & \textbf{73.35}
                  & \textbf{67.27}&\textbf{71.50}
                  & \textbf{61.35}
                  & \textbf{51.20}  \\
\bottomrule
\end{tabular}
\caption{Performance comparison on MMAU Test-mini, MMSU, and MMAR. For MMAU Test-mini, we report accuracy on the sound, music, and speech subsets, together with the overall average. Bold indicates the best result in each column among the listed open-source methods.}
\label{main_result}
\end{table*}

\subsection{AudioRE Dataset Analysis}
Figure~\ref{data_num}(a) summarizes the source and domain composition of AudioRE across the eight constituent datasets. The final corpus contains 286,300 instances covering sound, speech, and music.
Following prior work on structured audio reasoning~\cite{xie2025audio}, we further examine the distribution of CoT lengths in AudioRE, as shown in Figure~\ref{data_num}(b). AudioRE contains CoTs spanning a broad range of lengths, including both concise and more detailed answer explanations and thus providing varying amounts of textual reasoning supervision across instances.
Figure~\ref{after_tsne} presents a t-SNE visualization of the CLAP embeddings of the audio clips in AudioRE after acoustic redundancy filtering. Compared with the original aggregated data shown in Figure~\ref{fig:cross_dataset_similarity}(a), the retained samples are more evenly distributed across the embedding space with notably fewer high-density regions, qualitatively validating that the filtering strategy effectively mitigates acoustic redundancy while maintaining broad acoustic coverage. It should be noted that t-SNE is used here to illustrate the overall distribution trend of the samples, and the samples do not exhibit cluster-level structures based on their originating datasets.
The domain composition of AudioRE reflects the availability and annotation characteristics of existing public datasets. Therefore, AudioRE is intended as a general-purpose audio-language post-training corpus for LALMs.

\section{Experiments}\label{experiment}
\subsection{Experimental Setup}
\textbf{Training Details.}
We conduct full-parameter supervised fine-tuning of Qwen2-Audio-7B-Instruct~\cite{chu2024qwen2} on AudioRE to evaluate its effectiveness as a post-training resource. We train the model for two epochs using an initial learning rate of $1\times10^{-6}$ and a global batch size of 20. Checkpoints are saved every 100 optimizer steps. In the experiments, the optimal iteration result is selected for final analysis. Given our focus on audio understanding ability, CoT is excluded from the following evaluations. Further experimental details are provided in the supplementary material.

\textbf{Evaluation Metrics.}
We evaluate the models on multiple audio understanding and reasoning benchmarks using accuracy as the primary metric. MMAU Test-mini~\cite{sakshi2024mmau} serves as our main benchmark. It contains audio question-answering instances spanning sound, music, and speech, including questions that require multi-step reasoning. We report the accuracy for each domain as well as the average across the three domains. To evaluate performance beyond MMAU Test-mini, we additionally report results on MMSU~\cite{wang2026mmsu} and MMAR~\cite{ma2025mmar}.

\textbf{Baselines.}
We compare the model fine-tuned on AudioRE with representative LALMs, including LTU~\cite{gong2024listen}, LTU-AS~\cite{gong2023joint},  SALMONN~\cite{tang2024salmonn}, Qwen-Audio-Chat~\cite{chu2023qwen}, GAMA~\cite{ghosh2024gama}, GAMA-IT~\cite{ghosh2024gama}, Kimi-Audio~\cite{ding2025kimi},  and Mellow~\cite{deshmukh2025mellow}. We additionally include Qwen2-Audio-7B-Instruct under direct inference and SFT on other datasets, including Audio-Reasoner~\cite{xie2025audio}, R1-AQA~\cite{li2025reinforcement}, and SARI~\cite{wen2025sari}.

\begin{table*}[t]
\centering
\begin{tabular}{ccccc}
\toprule
\textbf{Dataset}
& \textbf{\# Instances}
& \textbf{MMAU Test-mini}
& \textbf{MMSU}
& \textbf{MMAR} \\
\midrule
Unfiltered aggregate& 592,810 & 69.40 & 59.33& 50.10 \\

Random subset (size-matched)& 286,300 & 70.00 & 58.43 & 50.40 \\

AudioRE& 286,300 & 71.50 & 61.35 & 51.20 \\
\bottomrule
\end{tabular}
\caption{
Comparison of different data selection strategies.
The random baseline is size-matched to AudioRE.
}
\label{tab:selection_ablation}
\end{table*}

\begin{figure*}[t]
\centering
\includegraphics[width=0.99\linewidth]{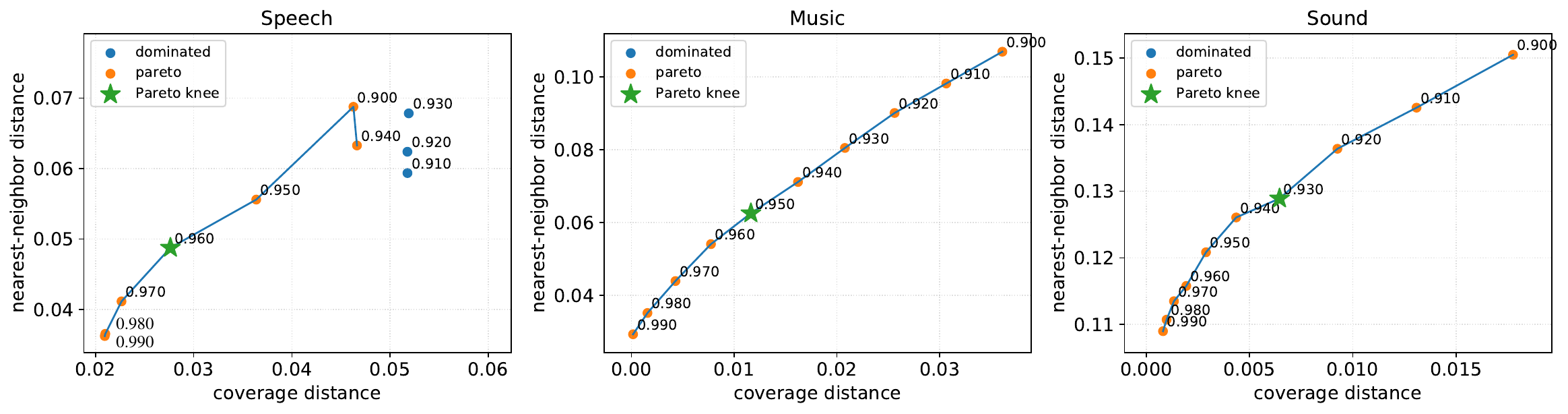}
\caption{
Domain-specific analysis of CLAP-based redundancy thresholds from 0.90 to 0.99. 
}
\label{fig:modality_threshold_selection}
\end{figure*}

\subsection{Main Results}
Table~\ref{main_result} reports the main results on MMAU Test-mini, MMSU, and MMAR. We first evaluate the effectiveness of AudioRE using Qwen2-Audio-7B-Instruct as the backbone model.
Compared with direct inference, Qwen2-AudioRE, SFT on AudioRE, consistently improves performance across the evaluated benchmarks. On MMAU Test-mini, the average accuracy increases from 59.60\% under direct inference to 71.50\%, corresponding to a performance improvement of 11.90\%. Consistent improvements are observed across all three audio domains: sound improves from 67.27\% to 73.87\%, music improves from 56.29\% to 73.35\%, and speech improves from 55.26\% to 67.27\%.
Qwen2-AudioRE also improves the performance of Qwen2-Audio-7B-Instruct from 35.72\% to 61.35\% on MMSU and from 30.00\% to 51.20\% on MMAR. 
The consistent gains across different benchmarks indicate that AudioRE provides effective supervision for improving audio understanding and reasoning. Compared with other Qwen2-Audio-7B-Instruct models fine-tuned on other datasets, Qwen2-AudioRE achieves consistently superior performance across the evaluated benchmarks.
In particular, Qwen2-AudioRE improves accuracy by 9.79\%, 12.15\%, and 14.40\% on MMAU Test-mini, MMSU, and MMAR compared with Audio-Reasoner, demonstrating the effectiveness of AudioRE as a post-training dataset.

\subsection{Selection of Acoustic Redundancy Thresholds}
Figure~\ref{fig:modality_threshold_selection} plots the coverage distance and nearest-neighbor distance associated with each candidate threshold.
We evaluate candidate redundancy thresholds from 0.90 to 0.99 at intervals of 0.01 for each audio domain. Each threshold is characterized by its coverage distance, which is minimized, and nearest-neighbor distance, which is maximized. After removing dominated candidates, we select the knee point on the normalized Pareto frontier using the maximum-distance-to-chord criterion. This yields $\tau_{\mathrm{sound}}^\star=0.93$, $\tau_{\mathrm{music}}^\star=0.95$, and
$\tau_{\mathrm{speech}}^\star=0.96$. The different knee points suggest that the three domains require different filtering strengths to balance coverage preservation and local redundancy.

\subsection{Effectiveness of Construction Pipeline}
To evaluate the effectiveness of data construction pipeline, we compare three training corpora: the unfiltered aggregate, a size-matched random subset, and AudioRE. The unfiltered aggregate contains 592,810 training instances, whereas AudioRE contains 286,300 instances, corresponding to only 48.30\% of the unfiltered corpus. The random subset contains 286,300 training instances, matching the size of AudioRE, and is sampled from the same candidate pool without using acoustic similarity. The random subset undergoes the same annotation construction and quality control procedures as AudioRE. All models are trained using the same Qwen2-Audio-7B-Instruct backbone, optimization settings, and number of epochs. As shown in Table~\ref{tab:selection_ablation}, SFT on the unfiltered aggregate achieves 69.40\%, 59.33\%, and 50.10\% on MMAU Test-mini, MMSU, and MMAR, respectively. Qwen2-AudioRE achieves 71.50\%, 61.35\%, and 51.20\% on the same benchmarks. These results show that Qwen2-AudioRE maintains comparable downstream performance with substantially fewer training instances. The random subset obtains 70.00\%, 58.43\%, and 50.40\% on MMAU Test-mini, MMSU, and MMAR, respectively. Compared with the random subset, Qwen2-AudioRE improves accuracy by 1.50\%, 2.92\%, and 0.80\% on the three benchmarks, respectively. The results suggest that redundancy-aware data selection provides more effective post-training supervision than size-matched random selection.

\section{Conclusion}\label{conclusion}
In this paper, we introduce AudioRE, a post-training dataset for LALMs covering the sound, speech, and music domains. AudioRE is constructed by integrating eight publicly available audio datasets into a unified supervision format. Its construction follows a three-stage pipeline consisting of acoustic redundancy filtering in the CLAP embedding space, unified MCQA and CoT construction, and a final quality verification and filtering. SFT experiments show that fine-tuning Qwen2-Audio-7B-Instruct on AudioRE consistently improves its performance on MMAU Test-mini, MMSU, and MMAR. These results indicate the practical value of AudioRE as a post-training resource and suggest that redundancy-aware data construction is a useful design for LALM post-training. 
Future work will expand the more domain coverage of AudioRE by incorporating a broader range of languages, speakers, acoustic conditions, and speech understanding tasks. 
AudioRE will facilitate future studies of data allocation across SFT and RL, cross-domain interactions during joint post-training, and domain composition under fixed data or computational budgets.

\bibliography{aaai2027}

@inproceedings{yang2022avqa,
  title={Avqa: A dataset for audio-visual question answering on videos},
  author={Yang, Pinci and Wang, Xin and Duan, Xuguang and Chen, Hong and Hou, Runze and Jin, Cong and Zhu, Wenwu},
  booktitle={Proceedings of the 30th ACM international conference on multimedia},
  pages={3480--3491},
  year={2022}
}

@article{tian2025audiox,
  title={Audiox: Diffusion transformer for anything-to-audio generation},
  author={Tian, Zeyue and Jin, Yizhu and Liu, Zhaoyang and Yuan, Ruibin and Tan, Xu and Chen, Qifeng and Xue, Wei and Guo, Yike},
  journal={arXiv preprint arXiv:2503.10522},
  year={2025}
}

@inproceedings{
bai2024audiosetcaps,
title={AudioSetCaps: Enriched Audio Captioning Dataset Generation Using Large Audio Language Models},
author={JISHENG BAI and Haohe Liu and Mou Wang and Dongyuan Shi and Wenwu Wang and Mark D Plumbley and Woon-Seng Gan and Jianfeng Chen},
booktitle={Audio Imagination: NeurIPS 2024 Workshop AI-Driven Speech, Music, and Sound Generation},
year={2024},
url={https://openreview.net/forum?id=uez4PMZwzP}
}

@inproceedings{tian2025vidmuse,
  title={Vidmuse: A simple video-to-music generation framework with long-short-term modeling},
  author={Tian, Zeyue and Liu, Zhaoyang and Yuan, Ruibin and Pan, Jiahao and Liu, Qifeng and Tan, Xu and Chen, Qifeng and Xue, Wei and Guo, Yike},
  booktitle={Proceedings of the Computer Vision and Pattern Recognition Conference},
  pages={18782--18793},
  year={2025}
}

@misc{chu2024qwen2,
      title={Qwen2-Audio Technical Report}, 
      author={Yunfei Chu and Jin Xu and Qian Yang and Haojie Wei and Xipin Wei and Zhifang Guo and Yichong Leng and Yuanjun Lv and Jinzheng He and Junyang Lin and Chang Zhou and Jingren Zhou},
      year={2024},
      eprint={2407.10759},
      archivePrefix={arXiv},
      primaryClass={eess.AS},
      url={https://arxiv.org/abs/2407.10759}, 
}

@misc{deshmukh2025mellow,
      title={Mellow: a small audio language model for reasoning}, 
      author={Soham Deshmukh and Satvik Dixit and Rita Singh and Bhiksha Raj},
      year={2025},
      eprint={2503.08540},
      archivePrefix={arXiv},
      primaryClass={cs.SD},
      url={https://arxiv.org/abs/2503.08540}, 
}

@misc{ding2025kimi,
      title={Kimi-Audio Technical Report}, 
      author={Ding Ding and Zeqian Ju and Yichong Leng and Songxiang Liu and Tong Liu and Zeyu Shang and Kai Shen and Wei Song and Xu Tan and Heyi Tang and Zhengtao Wang and Chu Wei and Yifei Xin and Xinran Xu and Jianwei Yu and Yutao Zhang and Xinyu Zhou and Y. Charles and Jun Chen and Yanru Chen and Yulun Du and Weiran He and Zhenxing Hu and Guokun Lai and Qingcheng Li and Yangyang Liu and Weidong Sun and Jianzhou Wang and Yuzhi Wang and Yuefeng Wu and Yuxin Wu and Dongchao Yang and Hao Yang and Ying Yang and Zhilin Yang and Aoxiong Yin and Ruibin Yuan and Yutong Zhang and Zaida Zhou},
      year={2025},
      eprint={2504.18425},
      archivePrefix={arXiv},
      primaryClass={eess.AS},
      url={https://arxiv.org/abs/2504.18425}, 
}

@misc{doh2023lp,
      title={LP-MusicCaps: LLM-Based Pseudo Music Captioning}, 
      author={SeungHeon Doh and Keunwoo Choi and Jongpil Lee and Juhan Nam},
      year={2023},
      eprint={2307.16372},
      archivePrefix={arXiv},
      primaryClass={cs.SD},
      url={https://arxiv.org/abs/2307.16372}, 
}

@inproceedings{kimAudioCaps2019,
    author= {Chris Dongjoo, Kim and Byeongchang, Kim and Hyunmin, Lee and Gunhee, Kim},
    title= {AudioCaps: Generating Captions for Audios in The Wild},
    booktitle = {NAACL-HLT},
    year = 2019
}

@misc{yang2025qwen3,
      title={Qwen3 Technical Report}, 
      author={An Yang and Anfeng Li and Baosong Yang and Beichen Zhang and Binyuan Hui and Bo Zheng and Bowen Yu and Chang Gao and Chengen Huang and Chenxu Lv and Chujie Zheng and Dayiheng Liu and Fan Zhou and Fei Huang and Feng Hu and Hao Ge and Haoran Wei and Huan Lin and Jialong Tang and Jian Yang and Jianhong Tu and Jianwei Zhang and Jianxin Yang and Jiaxi Yang and Jing Zhou and Jingren Zhou and Junyang Lin and Kai Dang and Keqin Bao and Kexin Yang and Le Yu and Lianghao Deng and Mei Li and Mingfeng Xue and Mingze Li and Pei Zhang and Peng Wang and Qin Zhu and Rui Men and Ruize Gao and Shixuan Liu and Shuang Luo and Tianhao Li and Tianyi Tang and Wenbiao Yin and Xingzhang Ren and Xinyu Wang and Xinyu Zhang and Xuancheng Ren and Yang Fan and Yang Su and Yichang Zhang and Yinger Zhang and Yu Wan and Yuqiong Liu and Zekun Wang and Zeyu Cui and Zhenru Zhang and Zhipeng Zhou and Zihan Qiu},
      year={2025},
      eprint={2505.09388},
      archivePrefix={arXiv},
      primaryClass={cs.CL},
      url={https://arxiv.org/abs/2505.09388}, 
}

@misc{agostinelli2023musiccaps,
      title={MusicLM: Generating Music From Text}, 
      author={Andrea Agostinelli and Timo I. Denk and Zalán Borsos and Jesse Engel and Mauro Verzetti and Antoine Caillon and Qingqing Huang and Aren Jansen and Adam Roberts and Marco Tagliasacchi and Matt Sharifi and Neil Zeghidour and Christian Frank},
      year={2023},
      eprint={2301.11325},
      archivePrefix={arXiv},
      primaryClass={cs.SD},
      url={https://arxiv.org/abs/2301.11325}, 
}

@inproceedings{ghosh2024gama,
  title={Gama: A large audio-language model with advanced audio understanding and complex reasoning abilities},
  author={Ghosh, Sreyan and Kumar, Sonal and Seth, Ashish and Evuru, Chandra Kiran Reddy and Tyagi, Utkarsh and Sakshi, S and Nieto, Oriol and Duraiswami, Ramani and Manocha, Dinesh},
  booktitle={Proceedings of the 2024 Conference on Empirical Methods in Natural Language Processing},
  pages={6288--6313},
  year={2024}
}

@misc{ghosh2025audio,
      title={Audio Flamingo 2: An Audio-Language Model with Long-Audio Understanding and Expert Reasoning Abilities}, 
      author={Sreyan Ghosh and Zhifeng Kong and Sonal Kumar and S Sakshi and Jaehyeon Kim and Wei Ping and Rafael Valle and Dinesh Manocha and Bryan Catanzaro},
      year={2025},
      eprint={2503.03983},
      archivePrefix={arXiv},
      primaryClass={cs.SD},
      url={https://arxiv.org/abs/2503.03983}, 
}

@misc{li2025reinforcement,
      title={Reinforcement Learning Outperforms Supervised Fine-Tuning: A Case Study on Audio Question Answering}, 
      author={Gang Li and Jizhong Liu and Heinrich Dinkel and Yadong Niu and Junbo Zhang and Jian Luan},
      year={2025},
      eprint={2503.11197},
      archivePrefix={arXiv},
      primaryClass={cs.SD},
      url={https://arxiv.org/abs/2503.11197}, 
}

@misc{ma2025audio,
      title={Audio-CoT: Exploring Chain-of-Thought Reasoning in Large Audio Language Model}, 
      author={Ziyang Ma and Zhuo Chen and Yuping Wang and Eng Siong Chng and Xie Chen},
      year={2025},
      eprint={2501.07246},
      archivePrefix={arXiv},
      primaryClass={cs.SD},
      url={https://arxiv.org/abs/2501.07246}, 
}

@misc{ma2025mmar,
      title={MMAR: A Challenging Benchmark for Deep Reasoning in Speech, Audio, Music, and Their Mix}, 
      author={Ziyang Ma and Yinghao Ma and Yanqiao Zhu and Chen Yang and Yi-Wen Chao and Ruiyang Xu and Wenxi Chen and Yuanzhe Chen and Zhuo Chen and Jian Cong and Kai Li and Keliang Li and Siyou Li and Xinfeng Li and Xiquan Li and Zheng Lian and Yuzhe Liang and Minghao Liu and Zhikang Niu and Tianrui Wang and Yuping Wang and Yuxuan Wang and Yihao Wu and Guanrou Yang and Jianwei Yu and Ruibin Yuan and Zhisheng Zheng and Ziya Zhou and Haina Zhu and Wei Xue and Emmanouil Benetos and Kai Yu and Eng-Siong Chng and Xie Chen},
      year={2025},
      eprint={2505.13032},
      archivePrefix={arXiv},
      primaryClass={cs.SD},
      url={https://arxiv.org/abs/2505.13032}, 
}

@misc{rouditchenko2025omni,
      title={Omni-R1: Do You Really Need Audio to Fine-Tune Your Audio LLM?}, 
      author={Andrew Rouditchenko and Saurabhchand Bhati and Edson Araujo and Samuel Thomas and Hilde Kuehne and Rogerio Feris and James Glass},
      year={2025},
      eprint={2505.09439},
      archivePrefix={arXiv},
      primaryClass={eess.AS},
      url={https://arxiv.org/abs/2505.09439}, 
}

@misc{sakshi2024mmau,
      title={MMAU: A Massive Multi-Task Audio Understanding and Reasoning Benchmark}, 
      author={S Sakshi and Utkarsh Tyagi and Sonal Kumar and Ashish Seth and Ramaneswaran Selvakumar and Oriol Nieto and Ramani Duraiswami and Sreyan Ghosh and Dinesh Manocha},
      year={2024},
      eprint={2410.19168},
      archivePrefix={arXiv},
      primaryClass={eess.AS},
      url={https://arxiv.org/abs/2410.19168}, 
}

@misc{wang2026mmsu,
      title={MMSU: A Massive Multi-task Spoken Language Understanding and Reasoning Benchmark}, 
      author={Dingdong Wang and Junan Li and Jincenzi Wu and Dongchao Yang and Xueyuan Chen and Tianhua Zhang and Helen Meng},
      year={2026},
      eprint={2506.04779},
      archivePrefix={arXiv},
      primaryClass={cs.CL},
      url={https://arxiv.org/abs/2506.04779}, 
}

@misc{xie2025audio,
      title={Audio-Reasoner: Improving Reasoning Capability in Large Audio Language Models}, 
      author={Zhifei Xie and Mingbao Lin and Zihang Liu and Pengcheng Wu and Shuicheng Yan and Chunyan Miao},
      year={2025},
      eprint={2503.02318},
      archivePrefix={arXiv},
      primaryClass={cs.SD},
      url={https://arxiv.org/abs/2503.02318}, 
}

@inproceedings{gong2023joint,
  title={Joint audio and speech understanding},
  author={Gong, Yuan and Liu, Alexander H and Luo, Hongyin and Karlinsky, Leonid and Glass, James},
  booktitle={2023 IEEE Automatic Speech Recognition and Understanding Workshop (ASRU)},
  pages={1--8},
  year={2023},
  organization={IEEE}
}

@misc{tang2024salmonn,
      title={SALMONN: Towards Generic Hearing Abilities for Large Language Models}, 
      author={Changli Tang and Wenyi Yu and Guangzhi Sun and Xianzhao Chen and Tian Tan and Wei Li and Lu Lu and Zejun Ma and Chao Zhang},
      year={2024},
      eprint={2310.13289},
      archivePrefix={arXiv},
      primaryClass={cs.SD},
      url={https://arxiv.org/abs/2310.13289}, 
}

@misc{goel2025audio,
      title={Audio Flamingo 3: Advancing Audio Intelligence with Fully Open Large Audio Language Models}, 
      author={Arushi Goel and Sreyan Ghosh and Jaehyeon Kim and Sonal Kumar and Zhifeng Kong and Sang-gil Lee and Chao-Han Huck Yang and Ramani Duraiswami and Dinesh Manocha and Rafael Valle and Bryan Catanzaro},
      year={2025},
      eprint={2507.08128},
      archivePrefix={arXiv},
      primaryClass={cs.SD},
      url={https://arxiv.org/abs/2507.08128}, 
}

@misc{xu2025qwen25omni,
      title={Qwen2.5-Omni Technical Report}, 
      author={Jin Xu and Zhifang Guo and Jinzheng He and Hangrui Hu and Ting He and Shuai Bai and Keqin Chen and Jialin Wang and Yang Fan and Kai Dang and Bin Zhang and Xiong Wang and Yunfei Chu and Junyang Lin},
      year={2025},
      eprint={2503.20215},
      archivePrefix={arXiv},
      primaryClass={cs.CL},
      url={https://arxiv.org/abs/2503.20215}, 
}

@misc{yang2026moss,
      title={MOSS-Audio Technical Report}, 
      author={Chen Yang and Chufan Yu and Hanfu Chen and Jie Zhu and Jingqi Chen and Ke Chen and Wenxuan Wang and Yang Wang and Yaozhou Jiang and Yi Jiang and Zhengyuan Lin and Ziqi Chen and Zhaoye Fei and Chenghao Liu and Donghua Yu and Jun Zhan and Kang Yu and Kexin Huang and Liwei Fan and Mingshu Chen and Qinyuan Cheng and Ruixiao Li and Shimin Li and Songlin Wang and Xingjian Zhao and Yang Gao and Yitian Gong and Yiyang Zhang and Zhe Xu and Xipeng Qiu},
      year={2026},
      eprint={2606.01802},
      archivePrefix={arXiv},
      primaryClass={cs.SD},
      url={https://arxiv.org/abs/2606.01802}, 
}

@misc{he2026measuring,
      title={Measuring Audio's Impact on Correctness: Audio-Contribution-Aware Post-Training of Large Audio Language Models}, 
      author={Haolin He and Xingjian Du and Renhe Sun and Zheqi Dai and Yujia Xiao and Mingru Yang and Jiayi Zhou and Xiquan Li and Zhengxi Liu and Zining Liang and Chunyat Wu and Qianhua He and Tan Lee and Xie Chen and Wei-Long Zheng and Weiqiang Wang and Mark Plumbley and Jian Liu and Qiuqiang Kong},
      year={2026},
      eprint={2509.21060},
      archivePrefix={arXiv},
      primaryClass={eess.AS},
      url={https://arxiv.org/abs/2509.21060}, 
}

@misc{wen2025sari,
      title={SARI: Structured Audio Reasoning via Curriculum-Guided Reinforcement Learning}, 
      author={Cheng Wen and Tingwei Guo and Shuaijiang Zhao and Wei Zou and Xiangang Li},
      year={2025},
      eprint={2504.15900},
      archivePrefix={arXiv},
      primaryClass={cs.CL},
      url={https://arxiv.org/abs/2504.15900}, 
}

@article{deshmukh2023pengi,
  title={Pengi: An audio language model for audio tasks},
  author={Deshmukh, Soham and Elizalde, Benjamin and Singh, Rita and Wang, Huaming},
  journal={Advances in Neural Information Processing Systems},
  volume={36},
  pages={18090--18108},
  year={2023}
}

@misc{défossez2024moshi,
      title={Moshi: a speech-text foundation model for real-time dialogue}, 
      author={Alexandre Défossez and Laurent Mazaré and Manu Orsini and Amélie Royer and Patrick Pérez and Hervé Jégou and Edouard Grave and Neil Zeghidour},
      year={2024},
      eprint={2410.00037},
      archivePrefix={arXiv},
      primaryClass={eess.AS},
      url={https://arxiv.org/abs/2410.00037}, 
}

@misc{laion_clap2022,
  doi = {10.48550/ARXIV.2211.06687},
  url = {https://arxiv.org/abs/2211.06687},
  author = {Wu, Yusong and Chen, Ke and Zhang, Tianyu and Hui, Yuchen and Berg-Kirkpatrick, Taylor and Dubnov, Shlomo},
  title = {Large-scale Contrastive Language-Audio Pretraining with Feature Fusion and Keyword-to-Caption Augmentation},
  publisher = {arXiv},
  year = {2022},
  copyright = {Creative Commons Attribution 4.0 International}
}

@inproceedings{drossos2020clotho,
  title={Clotho: An audio captioning dataset},
  author={Drossos, Konstantinos and Lipping, Samuel and Virtanen, Tuomas},
  booktitle={ICASSP 2020-2020 IEEE International Conference on Acoustics, Speech and Signal Processing (ICASSP)},
  pages={736--740},
  year={2020},
  organization={IEEE}
}

@misc{koizumi2023libritts,
      title={LibriTTS-R: A Restored Multi-Speaker Text-to-Speech Corpus}, 
      author={Yuma Koizumi and Heiga Zen and Shigeki Karita and Yifan Ding and Kohei Yatabe and Nobuyuki Morioka and Michiel Bacchiani and Yu Zhang and Wei Han and Ankur Bapna},
      year={2023},
      eprint={2305.18802},
      archivePrefix={arXiv},
      primaryClass={eess.AS},
      url={https://arxiv.org/abs/2305.18802}, 
}

@misc{kong2024audio,
      title={Audio Flamingo: A Novel Audio Language Model with Few-Shot Learning and Dialogue Abilities}, 
      author={Zhifeng Kong and Arushi Goel and Rohan Badlani and Wei Ping and Rafael Valle and Bryan Catanzaro},
      year={2024},
      eprint={2402.01831},
      archivePrefix={arXiv},
      primaryClass={cs.SD},
      url={https://arxiv.org/abs/2402.01831}, 
}

@inproceedings{melechovsky2024mustango,
  title={Mustango: Toward controllable text-to-music generation},
  author={Melechovsky, Jan and Guo, Zixun and Ghosal, Deepanway and Majumder, Navonil and Herremans, Dorien and Poria, Soujanya},
  booktitle={Proceedings of the 2024 Conference of the North American Chapter of the Association for Computational Linguistics: Human Language Technologies (Volume 1: Long Papers)},
  pages={8293--8316},
  year={2024}
}

@misc{ghosh2026audio,
      title={Audio Flamingo Next: Next-Generation Open Audio-Language Models for Speech, Sound, and Music}, 
      author={Sreyan Ghosh and Arushi Goel and Kaousheik Jayakumar and Lasha Koroshinadze and Nishit Anand and Zhifeng Kong and Siddharth Gururani and Sang-gil Lee and Jaehyeon Kim and Aya Aljafari and Chao-Han Huck Yang and Sungwon Kim and Ramani Duraiswami and Dinesh Manocha and Mohammad Shoeybi and Bryan Catanzaro and Ming-Yu Liu and Wei Ping},
      year={2026},
      eprint={2604.10905},
      archivePrefix={arXiv},
      primaryClass={cs.SD},
      url={https://arxiv.org/abs/2604.10905}, 
}

@article{wijngaard2025audio,
  author={Wijngaard, Gijs and Formisano, Elia and Esposito, Michele and Dumontier, Michel},
  journal={IEEE Access}, 
  title={Audio-Language Datasets of Scenes and Events: A Survey}, 
  year={2025},
  volume={13},
  number={},
  pages={20328-20360},
  doi={10.1109/ACCESS.2025.3534621}
}

@inproceedings{kim2019audiocaps,
  title={Audiocaps: Generating captions for audios in the wild},
  author={Kim, Chris Dongjoo and Kim, Byeongchang and Lee, Hyunmin and Kim, Gunhee},
  booktitle={Proceedings of the 2019 Conference of the North American Chapter of the Association for Computational Linguistics: Human Language Technologies, Volume 1 (Long and Short Papers)},
  pages={119--132},
  year={2019}
}

@misc{gong2024listen,
      title={Listen, Think, and Understand}, 
      author={Yuan Gong and Hongyin Luo and Alexander H. Liu and Leonid Karlinsky and James Glass},
      year={2024},
      eprint={2305.10790},
      archivePrefix={arXiv},
      primaryClass={eess.AS},
      url={https://arxiv.org/abs/2305.10790}, 
}

@misc{chu2023qwen,
      title={Qwen-Audio: Advancing Universal Audio Understanding via Unified Large-Scale Audio-Language Models}, 
      author={Yunfei Chu and Jin Xu and Xiaohuan Zhou and Qian Yang and Shiliang Zhang and Zhijie Yan and Chang Zhou and Jingren Zhou},
      year={2023},
      eprint={2311.07919},
      archivePrefix={arXiv},
      primaryClass={eess.AS},
      url={https://arxiv.org/abs/2311.07919}, 
}

\section{\Huge Supplementary Material}
\vspace{0.8cm}

\subsection{A. Source Dataset Details}
\label{sec:source_dataset_details}
AudioRE is constructed from eight publicly available audio datasets spanning three major domains: sound, music, and speech. For all source datasets, we use samples from their official training splits and exclude samples from the test splits. 

\paragraph{Sound Domain.}

For the sound domain, we collect four datasets: AVQA, Clotho, AudioCaps v2, and CompA-R.

\textit{AVQA.}
AVQA~\cite{yang2022avqa} is an audio-visual question-answering dataset collected from real-life scenarios. Its questions may rely on auditory evidence, visual evidence, or interactions between the two modalities, making it suitable for evaluating multimodal scene understanding and reasoning. Following~\cite{li2025reinforcement}, we replace ``video'' with ``audio'' where appropriate, while preserving the original answer choices and annotated correct answer. We collect 39,966 AVQA training instances as the initial candidate set. Acoustic redundancy filtering removes 5,272 instances, leaving 34,694 instances for quality verification. During quality verification, we remove questions that cannot be answered without visual information. We additionally discard samples containing inconsistencies among the caption, question, answer choices, annotated answer, and generated CoT. This stage removes a further 16,818 instances, resulting in 17,876 AVQA instances in the final AudioRE corpus.

\textit{Clotho.}
Clotho~\cite{drossos2020clotho} is a human-annotated, general-purpose audio-captioning dataset comprising audio clips collected from the Freesound platform. Each clip is accompanied by multiple natural-language captions describing the audible events and the overall acoustic scene. Clotho focuses on free-form descriptions of environmental sounds, everyday activities, and interactions among multiple sound events. We include 2,865 training instances from Clotho in the initial candidate set. Acoustic redundancy filtering removes 190 instances, leaving 2,675 Clotho instances in the final AudioRE corpus. 

\textit{AudioCaps v2.}
AudioCaps v2~\cite{kim2019audiocaps} is a large-scale, in-the-wild audio-captioning dataset constructed from AudioSet clips sourced from YouTube. Its captions are written via crowdsourcing on Amazon Mechanical Turk. We use the AudioCaps v2 release and collect 90,549 training instances for the initial candidate set. Our acoustic redundancy analysis identifies substantial overlap between some AudioCaps v2 clips and other YouTube- and AudioSet-derived corpora included in AudioRE. Acoustic redundancy filtering removes 11,328 instances. All retained instances pass the subsequent quality-verification stage and are included in the final AudioRE corpus, leaving 79,221 AudioCaps v2 instances.

\textit{CompA-R.}
CompA-R~\cite{ghosh2024gama} is a pipeline-generated instruction-tuning dataset. It is constructed from the AudioSet-strong subset and uses multiple forms of audio-related metadata to generate instruction-response pairs. Unlike conventional audio-captioning
datasets, its instructions often require reasoning about relationships among acoustic events, their likely contexts, and plausible acoustic events. We include 197,218 CompA-R training instances in the initial candidate set. Our acoustic redundancy analysis reveals substantial overlap between CompA-R and other AudioSet-derived
corpora, together with considerable redundancy within CompA-R itself. Acoustic redundancy filtering therefore removes 141,758 instances. All retained instances pass the subsequent quality-verification stage and are included in the final AudioRE corpus, leaving 55,460 CompA-R instances.

\paragraph{Music Domain.}

The music-domain datasets are smaller than the sound and speech datasets. Consequently, the number of removed music instances is lower than that of large datasets such as CompA-R and LibriTTS-R. 

\textit{MusicCaps.}
MusicCaps~\cite{agostinelli2023musiccaps} is an expert-annotated music-text dataset released with MusicLM. Each example consists of a 10-second music clip extracted from YouTube and a detailed natural-language description written by music experts. The annotations describe a broad range of perceptual and semantic properties, including genre, mood, tempo, singer voices, instrumentation, dissonances, and rhythm. We include 2,608 MusicCaps training instances in the initial source dataset. Acoustic redundancy filtering removes 47 instances, leaving 2,561 MusicCaps instances in the final AudioRE corpus. 

\textit{MusicBench.}
MusicBench~\cite{melechovsky2024mustango} is introduced for text-to-music generation. Its captions combine general semantic descriptions with music-specific information
obtained using music-information-retrieval tools. In addition to describing the overall musical content, the captions may include explicit attributes related to chords, beats, key, and tempo. These annotations make MusicBench useful for learning fine-grained musical attributes and music understanding. We collect 17,861 MusicBench instances in the initial candidate set. Acoustic redundancy filtering removes 8,614 instances, leaving 9,247 instances for subsequent quality verification. The quality-verification stage removes an additional 76 instances, resulting in 9,171 MusicBench instances in the final AudioRE corpus. 

\textit{LP-MusicCaps-MTT.}
LP-MusicCaps~\cite{doh2023lp} is a large-scale pseudo music caption dataset for music, constructed by transforming structured music tags into description sentences using a large language model. In this work, we use the MagnaTagATune subset, denoted as LP-MusicCaps-MTT. The original MagnaTagATune annotations describe attributes such as genre, instrumentation, vocals, mood, perceptual tempo, origin, and acoustic characteristics. The LP-MusicCaps pipeline verbalizes these structured tags to produce natural-language music descriptions with greater linguistic diversity. We collect 15,560 LP-MusicCaps-MTT instances in the initial candidate set. Acoustic redundancy filtering removes 4,781 instances, leaving 10,779 instances in the final AudioRE corpus.

\paragraph{Speech Domain.}

The speech domain is represented primarily by LibriTTS-R. Some sound-domain datasets, particularly AVQA and CompA-R, also contain audio clips involving speech. However, we do not separately extract or relabel these clips as speech instances. Instead, each sample retains the domain assignment of its source dataset. Therefore, the speech-domain statistics reported in this work refer specifically to the LibriTTS-R dataset.

\textit{LibriTTS-R.}
LibriTTS-R~\cite{koizumi2023libritts} is a restored multi-speaker speech corpus designed primarily for text-to-speech research. It is re-derived from the original materials (MP3 from LibriVox and texts from Project Gutenberg) of LibriSpeech. Each speech segment is paired with its corresponding textual transcription. LibriTTS-R preserves the speaker and utterance organization of LibriTTS while applying speech-restoration techniques to improve recording quality. We first collect 226,183 LibriTTS-R training segments. The speech segments and their associated textual annotations are aligned with existing question-answer annotations and normalized into the unified MCQA format used by AudioRE. Acoustic redundancy filtering removes 117,626 segments. All retained instances pass the subsequent quality-verification stage and are included in the final AudioRE corpus, leaving 108,557 instances.

\begin{table*}[t]
\centering
\small
\caption{Detailed statistics of the source datasets used to
construct AudioRE.}
\label{tab:source_stage_statistics}
\begin{tabular}{llccr}
\toprule
\textbf{Dataset}
& \textbf{Domain}
& \textbf{Candidates}
& \textbf{After Acoustic Filtering}
& \textbf{Final} \\
\midrule
AVQA              & Sound  & 39,966  & 34,694  & 17,876 \\
Clotho            & Sound  & 2,865   & 2,675   & 2,675 \\
AudioCaps v2      & Sound  & 90,549  & 79,221  & 79,221 \\
CompA-R           & Sound  & 197,218 & 55,460  & 55,460 \\
MusicCaps         & Music  & 2,608   & 2,561   & 2,561 \\
MusicBench        & Music  & 17,861  & 9,247   & 9,171 \\
LP-MusicCaps-MTT  & Music  & 15,560  & 10,779  & 10,779 \\
LibriTTS-R        & Speech & 226,183 & 108,557 & 108,557 \\
\midrule
\textbf{Total}
& --
& \textbf{592,810}
& \textbf{303,194}
& \textbf{286,300} \\
\bottomrule
\end{tabular}
\end{table*}

Acoustic redundancy filtering reduces the candidate corpus from 592,810 to 303,194 instances. The quality verification further removes 16,894 instances, resulting in 286,300 final training instances. The overall retention rate relative to the initial candidate corpus is 48.30\%. Six of the eight source datasets do not provide native QA annotations. Since our primary objective is to remove acoustically redundant samples during dataset construction, thereby improving data efficiency and mitigating diminishing returns in downstream performance, we aim to preserve high-quality existing annotations whenever possible. We therefore collect QA pairs from AudioMCQ~\cite{he2026measuring} and audio captions from AudioX-IFcaps~\cite{tian2025audiox,tian2025vidmuse} and AudioSetCaps~\cite{bai2024audiosetcaps}. In the final release of AudioRE, we will explicitly document the provenance of all audio sources.

\subsection{B. Chain-of-Thought Generation Prompt}
\label{sec:cot_prompt}

After converting the retained samples into the unified MCQA format, we use Qwen3-30B to generate a reasoning trace for each instance. Given an audio caption $c_i$, a question $q_i$, four answer choices $O_i$, and the annotated correct answer $y_i$, the generated reasoning is required to remain grounded in the acoustic evidence described by the caption. Each reasoning trace contains two components. The first component analyzes the question and identifies the type of acoustic evidence needed to answer it. In our implementation, the question type is determined by the domain category of the source dataset, namely sound, music, or speech. The second component connects the relevant evidence in the audio caption to the annotated answer and briefly distinguishes it from the distractors. Both components are constrained to be concise so that the resulting reasoning supervision does not dominate the original question and answer content. 

The exact prompt template is provided below.

\begin{promptbox}
TASK: Complete the THINKING PROCESS for this audio-based multiple-choice question using clear Chain-of-Thought reasoning.

QUESTION DETAILS:
- Question: ***question_text***
- Question Type: ***question_type***
- Choices: ***multi_choice***
- Answer: ***answer***

INCOMPLETE THINKING PROCESS:
According to the question text, <first_analysis>...</first_analysis>, so the question type is ***question_type***. I need to first analyze the audio content: ***caption***. According to the audio content, <second_analysis>...</second_analysis>, so the correct answer is ***answer***.

COMPLETION REQUIREMENTS:

FIRST THINKING PROCESS (Question Analysis):
- Start with a lowercase letter
- One continuous paragraph, no breaks
- Be analytical and methodical
- Maintain coherence with context
- Must: identify what the question asks, specify needed audio evidence, link to question type

SECOND THINKING PROCESS (Audio Analysis & Answer Selection):
- Start with a lowercase letter
- One continuous paragraph, no breaks
- Maintain coherence with context
- Use "quotes" only for choice options, never for audio
- Must: highlight key evidence from audio, explain how it supports the correct answer, briefly show why other options don't fit, connect evidence to answer

OUTPUT FORMAT:
- Place first process in <first_analysis> </first_analysis>
- Place second process in <second_analysis> </second_analysis>
- Output nothing else.
\end{promptbox}

\subsection{C. Quality Verification and Filtering Prompt}
\label{sec:quality_filtering_prompt}
After MCQA normalization and CoT generation, we apply a quality-verification procedure to all instances before including them in the final AudioRE corpus. For each candidate instance, the review model receives the audio caption, question, four choices, correct answer, and generated CoT. The model determines whether these fields form a clear, consistent, and valid multiple-choice training instance.

Specifically, the review procedure checks the following criteria:
\begin{enumerate}
    \item the question is grammatically clear and semantically complete;
    
    \item the question can be answered using the acoustic evidence described by the caption;
    
    \item the four choices are distinguishable, and exactly one choice is correct;
    
    \item the answer is supported by the caption and is consistent with the question;
    
    \item the generated CoT is logically coherent and does not contain hallucinated or unsupported information;
    
    \item the caption, question, choices, annotated answer, and CoT are mutually consistent.
\end{enumerate}

For AVQA instances, we additionally reject questions that depend on visual-only evidence, including questions whose answers cannot be inferred from the available audio evidence. In addition, we conducted a manual spot-check on the filtered data. Batches of data undergo quality assurance via manual review by one human annotator.

The exact quality-verification prompt is shown below.

\begin{promptbox}
You are a data reviewer. You will be given an audio clip's information together with its caption, question, choices_text, answer, and the CoT used to solve the task. Your job is to verify whether these elements are consistent and supported by the acoustic evidence described in the caption.

Pay special attention to the reasoning process and check whether it contains any hallucinations, unsupported claims, or logical inconsistencies. Also verify whether the caption, question, choices, answer, and CoT are internally consistent.

Return <True> if everything is correct and there are no issues.
Return <False> if you find any error, inconsistency, or hallucination.

Here is the caption of the audio: "{caption}".
Here is the question: "{question}".
Here are the choices: "{choices_text}".
Here is the answer: "{answer}".
Here is the reasoning process: "{CoT}".

Please perform a careful and thorough evaluation and return the result in the specified format only.
Output exactly one label.
No explanation.
No extra text.

Output:
<True> or <False>
\end{promptbox}

\subsection{D. Qualitative Examples}

\begin{examplebox}{\textbf{Example 1: AVQA}}

\textit{Question:} Where did the audio take place?

\textit{A:} In the mountains

\textit{B:} On the railway

\textit{C:} On the road

\textit{D:} In the park

\textit{Qwen3-30B CoT:}
According to the question text, the question requires location identification from audio cues so evidence must match environmental sounds described in the audio without prior content reference, so the question type is "sound". I need to firstly analyze the audio content: This audio contains one instance of bus sounds, and traffic noise sounds. According to the audio content, bus and traffic noise evidence directly indicates urban transportation setting eliminating mountains and park; railway implies train sounds not present so "on the road" is correct, so the correct answer is "On the road".
\end{examplebox}

\begin{examplebox}{\textbf{Example 2: Clotho}}

\textit{Question:} What specific environmental sound event occurs in the audio?

\textit{A:} Leaves and drizzle in a forest 

\textit{B:} Waves and breeze near a shore 

\textit{C:} Thunder and hail striking a roof

\textit{D:} Wind and rain striking a surface

\textit{Qwen3-30B CoT:}
According to the question text, the question asks for a specific environmental sound event requiring identification through auditory cues the evidence needed involves recognizing distinct natural sounds and how they interact with their environment aligning with the sound question type, so the question type is "sound". I need to firstly analyze the audio content: A storm produces heavy winds and light rain which hit a window. According to the audio content, the audio features heavy winds and light rain hitting a surface which directly matches "Wind and rain striking a surface" while "Leaves and drizzle", "Waves and breeze near a shore" and "Thunder and hail striking a roof" are inconsistent with the described storm scenario, so the correct answer is "Wind and rain striking a surface".
\end{examplebox}

\begin{examplebox}{\textbf{Example 3: AudioCaps v2}}

\textit{Question:} What happens first in the audio?

\textit{A:} A bicycle bell rings twice

\textit{B:} A truck engine starts and moves away

\textit{C:} A car engine revs loudly

\textit{D:} A motorcycle engine idles then stops

\textit{Qwen3-30B CoT:}
According to the question text, the question asks what happens first in the audio and requires identifying the initial sound event from the given choices which must be determined through careful listening and analysis of the audio content, so the question type is "sound". I need to firstly analyze the audio content: A motorcycle engine idles then turns off and a man speaks. According to the audio content, the audio begins with a motorcycle engine idling then turning off which directly matches the description in "A motorcycle engine idles then stops" while the other choices involving a bicycle bell, a truck or a loud car engine revving are not supported by the actual audio content, so the correct answer is "A motorcycle engine idles then stops".
\end{examplebox}

\begin{examplebox}{\textbf{Example 4: CompA-R}}

\textit{Question:} Based on the audio, what is the most likely reason for the loud engine sounds and crowd cheers?

\textit{A:} A boat race is happening on water

\textit{B:} A construction truck is working on site

\textit{C:} A race car is performing in a stadium

\textit{D:} A plane is taking off nearby

\textit{Qwen3-30B CoT:}
According to the question text, the question asks for the most likely reason for loud engine sounds and crowd cheers, requiring identification of the source event through audio cues, linking auditory evidence to the correct scenario among the given choices, so the question type is sound. I need to firstly analyze the audio content: The roaring engine and cheers imply it could be a race car or a stunt vehicle performing in a stadium or on a race track. According to the audio content, the consistent engine roar and energetic cheers best match a live motorsport event; "A race car is performing in a stadium" fits this context, while boat race, construction truck, and plane sounds would differ significantly, so the correct answer is "A race car is performing in a stadium".
\end{examplebox}

\begin{examplebox}{\textbf{Example 5: MusicCaps}}

\textit{Question:} Which audio detail is most prominent in this live performance recording?

\textit{A:} Clear female vocal harmony

\textit{B:} Harsh microphone feedback sound

\textit{C:} Fast electronic drumbeat

\textit{D:} Bright piano melody line

\textit{Qwen3-30B CoT:}
According to the question text, the question asks which audio detail is most prominent in the live performance recording, requiring identification of the most noticeable sonic element. As a music question, it necessitates careful listening to distinguish between the presented options based on auditory prominence, so the question type is "music". I need to firstly analyze the audio content: The low quality recording features a live performance of a traditional song and it consists of flat male vocal and reverberant fruity male vocal singing over smooth bass and strings melody. It sounds very muddy, muffled and there is a harsh microphone feedback sound. The recording is noisy itself. According to the audio content, the harsh microphone feedback sound is clearly present and stands out amid the muffled and noisy recording. While male vocals, bass, and strings are audible, they are not among the listed choices. "Clear female vocal harmony", "Fast electronic drumbeat", and "Bright piano melody line" are absent, making "Harsh microphone feedback sound" the only accurate option, so the correct answer is "Harsh microphone feedback sound".
\end{examplebox}

\begin{examplebox}{\textbf{Example 6: LP-MusicCaps-MTT}}

\textit{Question:} What is the most prominent instrument heard in this piece?

\textit{A:} The violin

\textit{B:} The piano

\textit{C:} The flute

\textit{D:} The trumpet

\textit{Qwen3-30B CoT:}
According to the question text, the question asks for the most prominent instrument in the audio, which requires identifying the main instrument heard. As a music question, it necessitates analyzing the instrumentation of the provided piece without relying on prior knowledge or external assumptions, so the question type is "music". I need to firstly analyze the audio content: A beautiful classical piece filled with strings and the enchanting sound of the violin. According to the audio content, the audio features a classical piece dominated by strings and a standout violin melody described as enchanting, which aligns with "The violin". Other options like "The piano", "The flute", and "The trumpet" are not mentioned in the description, making "The violin" the most prominent, so the correct answer is "The violin".

\end{examplebox}

\begin{examplebox}{\textbf{Example 7: MusicBench}}

\textit{Question:} What effect is applied to the vocals in the audio to create a feeling of anticipation?

\textit{A:} Auto-tune

\textit{B:} Echo

\textit{C:} Backwards reverb

\textit{D:} Flanger

\textit{Qwen3-30B CoT:}
According to the question text, the question asks for a vocal effect creating anticipation in music requiring audio evidence to identify the correct choice from the given options without referencing specific song details, so the question type is "music". I need to firstly analyze the audio content: Listen to this educational song where a male singer seductively croons over a beat of three. Through the use of a Backwards reverb effect on the vocals, the track creates a sense of anticipation. This technique would be a great example for any sound design course. For those interested in music, this song showcases the complexity of sound design and how it can be used to elevate a track. So sit back, listen closely, and enjoy this instructive recording. According to the audio content, the audio explicitly states "Backwards reverb effect" creates anticipation, while "Auto-tune", "Echo", and "Flanger" are not referenced as the applied technique for this effect, so the correct answer is "Backwards reverb".
\end{examplebox}

\begin{examplebox}{\textbf{Example 8: LibriTTS-R}}

\textit{Question:} What emotional delivery does the speaker use?

\textit{A:} Calm and mature tone

\textit{B:} Monotone and disinterested delivery

\textit{C:} Energetic and youthful tone

\textit{D:} Angry and forceful delivery

\textit{Qwen3-30B CoT:}
According to the question text, the question asks about the speaker's emotional delivery, requiring analysis of tone and expression in the audio. Since the question type is speech, the focus is on vocal characteristics rather than content, so the correct choice must align with audible cues indicating emotion, so the question type is "speech". I need to firstly analyze the audio content: In a fast-paced and high-pitched voice, a youth with natural emotions states, Also I am going to start among our older children a limited form of self-government such as we had in college. According to the audio content, the speaker uses a "Fast-paced and high-pitched voice" and is described as a "Youth with natural emotions", directly pointing to an "Energetic and youthful tone". Other options like "Calm and mature" or "Monotone" contradict the liveliness in the delivery, while "Angry and forceful" lacks supporting evidence, so the correct answer is "Energetic and youthful tone".
\end{examplebox}

\subsection{E. Experimental Parameters}
\label{sec:experimental_parameters}

\paragraph{Training Setting.}

We conduct full-parameter supervised fine-tuning of Qwen2-Audio-7B-Instruct for two epochs. Before being fed into the model, all audio waveforms are resampled to 16~kHz. The maximum sequence length is set to 1,024. We use a cosine learning-rate scheduler. The learning rate is linearly warmed up over the first 5\% of the total training steps, which corresponds to a warm-up ratio of 0.05. Table~\ref{tab:sft_hyperparameters} summarizes the complete training configuration.

\begin{table}[h]
\centering
\small
\begin{tabular}{lc}
\toprule
\textbf{Hyperparameter} & \textbf{Value} \\
\midrule
Training strategy
    & Full-parameter SFT \\
Number of training epochs
    & 2 \\
Maximum sequence length
    & 1,024  \\
Effective batch size
    & 20 \\
Gradient accumulation steps
    & 1 \\
Initial learning rate
    & $1\times10^{-6}$ \\
Learning-rate scheduler
    & Cosine \\
Warm-up ratio
    & 0.05 \\
Weight decay
    & 0.01 \\
Audio sampling rate
    & 16~kHz \\
Data type & bfloat16  \\
Distributed training & DeepSpeed ZeRO-3 \\
\bottomrule
\end{tabular}
\caption{
Training configurations used in supervised fine-tuning experiments.}
\label{tab:sft_hyperparameters}
\end{table}

\paragraph{Evaluation prompt.}

For benchmark evaluation, we use a unified prompt template. The model receives the audio clip, question, and four answer choices and is instructed to return only the selected answer enclosed by the \texttt{<answer>} and \texttt{</answer>} tags. Neither the source caption nor the generated reasoning trace is included in the evaluation input.

In the following template, \texttt{|<Question>|} and \texttt{|<Choices>|} are replaced with the benchmark question and its corresponding answer choices, respectively.

\begin{promptbox}
|<Question>| Please choose the answer from the following options: |<Choices>|. Output the final answer in <answer> </answer>.
\end{promptbox}


\end{document}